# EFFECT OF STRAIN ON THE TRANSPORT PROPERTIES OF THE MANGANITE SYSTEMS.

S. Basak , I. Chaudhuri *  and S. K. Ghatak

**Department of Physics & Meteorology, Indian Institute of Technology,
Kharagpur - 721302, India.**

**ABSTRACT:**

The effect of strain on the resistivity and thermopower of ferromagnetic manganites has been examined based on the model that incorporates the electron-lattice interaction through the Jahn-Teller effect and an effective hopping determined by nearest neighbour spin-spin correlation of $t_{2g}$ electrons. The metal insulator transition temperature associated with resistivity decreases with increase in strain. In the presence of large strain the system remains in the semiconducting state. Thermopower (S) is positive and increasing function of strain and it exhibits a maximum with temperature. The temperature where maximum of S appears, shifts towards higher (lower) value with in the presence of magnetic field (strain). A large magneto-thermopower that depends on strain is obtained around metal-insulator transition.

* indira@phy.iitkgp.ernet.in

## Introduction:

Recently there has been extensive studies on the properties of hole-doped manganese oxides e.g. $R_{1-x}A_xMnO_3$ (R = rare-earth, A = Ca, Sr) with perovskite structure, arising from their many unusual physical properties and rich phase diagram[1,2]. For 0.2<x<0.5 these compounds undergo a phase transition from paramagnetic-insulator to ferromagnetic - metal as it is cooled down through the ferromagnetic transition temperature and exhibits large magnetoresistance. The transport properties in these systems are considered to be dominated by double exchange interaction between $Mn^{3+}$ and $Mn^{4+}$ ions[3-6]. The essence of the electronic transport in the double exchange model is that when an electron in an itinerant state hops from one site to its neighbouring site, the hopping is facilitated if it keeps its spin parallel to the spin S of localized electrons. In this process the Hund's rule energy (coupling) remains the same whereas the energy is gained due to delocalization. This process leads to an effective hopping integral that becomes a function of the spin-spin correlation of localized electrons[4]. It has been argued that the electron-phonon interaction also plays a very dominant role in these oxides. This notion is further strengthened by the observation that the transport properties of these oxides depend strongly on the average ionic size of the rare-earth $R^{7,8}$. The metal- insulator transition temperature decreases with decreasing average ionic radius of the R-ion and the system becomes more resistive. The important effect arising from a smaller R-ion is to induce lattice distortion. Many novel features of these oxides are the result of several competing interactions that originate from coupling of spin-charge and lattice degrees of freedom. The

lattice–charge coupling arises through the Jahn-Teller splitting of the $e_g$ - level of Mn and the double exchange interaction provides spin-charge coupling[9]. The importance of the Jahn-Teller effect in manganites is borne out by experimental evidence[10]. A model theoretical study demonstrates that the ferromagnetism and the Jahn-Teller distortion interfere destructively in the $e_g$ –band[11] and therefore it can change the resistive behaviour. In this work we explore the influence of the Jahn-Teller splitting due to distortion that is induced by average size reduction of the R-ion on electronic transport properties.

## Model and Calculations:

We consider a simple model where the mobile charge carriers are in an $e_g$ -state and the localized electrons in a $t_{2g}$ – orbital, is half-filled for possible configuration of Mn-ion in manganite. The localized electrons give rise to a spin S = 3/2 in ground state which is coupled to the

$e_g$ - electron through strong Hund's rule coupling. The orbital degeneracy of the $e_g$ -level is assumed to be lifted by the Jahn-Teller (J-T) distortion resulting from the chemical pressure due to the substitution of rare-earth ions with lower atomic diameter. The model can be described by a Hamiltonian:

$$H = \sum_{i\alpha\sigma} E_{d\alpha} d^+_{\alpha i\sigma} d_{\alpha i\sigma} + \sum_{ij\sigma,\alpha\neq\alpha'} t_{ij} d^+_{\alpha i\sigma} d_{\alpha' j\sigma} + J\sum_i S_i . \sigma_i \qquad (1)$$

The first term describes the site energy for $e_g$ electrons with energy $E_{d\alpha}$ and $d^+_{\alpha i\sigma}$ the creation operator for states $\alpha$ =1 and 2 at site i with spin $\sigma$. Due to the J-T effect, the site energy of two $e_g$ -states becomes $E_{d1,2} = E_d^o - (-1)^\alpha Ge$ where G is the electron - lattice interaction strength and e being the strain produced due to mismatch of ionic radii of R-ions. The second term in equation (1) represents the hopping between $e_g$ –levels of neighbouring sites i and j sites. The intra- and inter- orbital hopping integrals are taken to be the same. The last term is the exchange (Hund's rule) coupling between spin $S_i$ of $t_{2g}$- and $\sigma_i$ of $e_g$ -electrons. (///) It was shown that the strong exchange coupling modifies the hopping ($t_{ij}$) of $e_g$ electron in such a way that it depends on the spin configuration of neighbouring sites and its effect on charge transport which can be taken into account by considering $t_{ij} = t_{dd} (1 + <S_i.S_j>/ S^2)^{1/2}$ [9] where $<S_i.S_j>$ is the spin-spin correlation function. It is evident that the band-width of the $e_g$ band becomes temperature dependent through the spin-spin correlation function that is augmented as ferromagnetic order sets in. Eqation 1. is then reduced to that of two-band system with an effective nearest -neighbour hopping integral

$t = t_{dd} (1 + <S_i.S_j>/ S^2)^{1/2}$ . The transport properties are then obtained from the current-current response function. $\Pi_{xx} (\tau) = - << T_\tau (j_x (\tau), j_x(0))>>$, the brackets $(< >)$ define the thermal average, $T_\tau$ is the time-ordering operator and the longitudinal current density $j_x$ is given by

$$j_x = e \sum_{k\sigma} (\delta \varepsilon_{dk} / \delta k_x )[ (d^+_{1k\sigma}d_{1k\sigma}+d^+_{2k\sigma}d_{2k\sigma})+ (d^+_{1k\sigma}d_{2k\sigma} + d^+_{2k\sigma}d_{1k\sigma})]$$

with $\varepsilon_{dk} = 2t [ \cos (k_x a) + \cos (k_y a) + \cos (k_z a)]$. According to the Kubo formula[12], the frequency dependent conductivity can be expressed as

$$\sigma_{xx} (\omega) = -(1/\omega \Omega) \, \text{Im} \, \Pi_{xx} (\omega) \qquad (2)$$

where $\Pi_{xx}(\mathbf{q}, i\omega) = \int_0^\beta e^{i\omega\tau} \Pi_{xx}(\mathbf{q},\tau)\, d\tau$ is the Fourier transform of the current-current response function, $\Omega$ is the cell volume and $\beta = 1/kT$. The d.c. conductivity, defined as the zero frequency limit of the frequency dependent conductivity $\sigma(\omega)$, is then given by

$$\sigma_{dc} = \int (\partial f_o / \partial \varepsilon)\, L(\varepsilon)\, d\varepsilon \tag{3}$$

where $f_o = 1/[\exp\{(\varepsilon-\mu)\beta\} + 1]$, the Fermi function and the conductivity function $L(\varepsilon)$ is given by,

$$L(\varepsilon) = e^2 \sum_{k\sigma} (\delta\varepsilon_{dk}/\delta k_x)^2 [A_{11}^2 + A_{22}^2 + 4A_{21}^2 + 4A_{12}A_{11} + 2A_{11}A_{22} + 4A_{22}A_{12}] \tag{4}$$

The spectral weight functions $A_{\alpha\beta}(\varepsilon)$ are defined in terms of a one-electron Green's functions $G_{\alpha\beta}(k,\sigma,\varepsilon) = \langle\langle T_\tau(d_{\alpha k\sigma}; d^+_{\beta k\sigma})\rangle\rangle$ as

$$A_{\alpha\beta} = A_{\alpha\beta}(k, \varepsilon) = -1/\pi\, \mathrm{Im}\, G_{\alpha\beta}(k,\sigma,\varepsilon) \tag{5}$$

The chemical potential $\mu$ is obtained from the relation that the electron concentration (n) in an $e_g$-state is given by

$$n = \int \rho_d(\varepsilon)\, f_o(\varepsilon)\, d\varepsilon \tag{6}$$

where $\rho_d(\varepsilon) = \rho_{d1}(\varepsilon) + \rho_{d2}(\varepsilon)$, the density of states for $e_g$- orbitals.

The thermopower (S) is given by [12]

$$S = -\{1/e\sigma_{dc}T)\} \int (\partial f_o/\partial\varepsilon)(\varepsilon-\mu)\, L(\varepsilon)\, d\varepsilon \tag{7}$$

The spin correlation function of $t_{2g}$ - electrons is calculated by assuming nearest neighbour Heisenberg interaction between the spins $S_i$ and $S_j$ and using the cluster-variational method[15]. In this method the system is assumed to be built out of building blocks where interaction between spins within the block are treated exactly and the rest of the interactions are replaced by an effective field which is treated as a variational parameter. Assuming a pair of spins constitutes the building block, the spin correlation function $<S_i.S_j>$ and the magnetization are obtained as a function of temperature (Fig –1 inset). The critical temperature $T_c$ is determined by an effective exchange interaction between spins (S) of $t_{2g}$ -electrons.

### Results and Discussions:

The resistivity and thermo-electric power of the manganite system have been calculated numerically for different values of strain e. The parameters are normalized in terms of bare bandwidth ($12t_{dd}$) which is taken to be 2 eV and the energy is measured with reference $E_d^o = 0$. The strain is parameterized in term of the energy separation (Ge) between $e_g$ - orbitals and the results are presented for Ge = 0.05, 0.15, 0.25 and 0.35 and hole concentration = 0.2 per Mn-site. The temperature dependence of resistivity (ρ) of the manganite system for different values of strain is shown in Fig. 1. The resisitivity slowly increases as the temperature is lowered in the paramagnetic phase and drops sharply below the para- to ferro- magnetic transition temperature $T_c$. For $T<T_c$, the state is metallic –like with dρ/dT positive. The extent of the drop in resistivity depends sensitively on distortion. For large strain the system remains semi-conducting at all temperatures. The $e_g$-bandwidth is enhanced due to spin-spin correlation that is large at low temperatures ($T \ll T_c$) and decreases with increase in T (inset of Fig.1). Above $T_c$, the widening of the band-width persists arising from the short range spin-spin correlation, and this makes the system resistive compared to the situation where the spin-spin correlation is treated in mean field approximation. The $e_g$-. The transition from semiconducting to metallic transition around $T_c$ is due to the increase in the carrier mobility that results from large spin-spin correlation. The transition is also marked by a sharp change of the chemical potential around the transition temperature. For a less strained system, the density of states exhibits a two-peak structure where peaks appear around energy ≈ ± Ge/2 with a minimum at $E_d^0 = 0$ and the Fermi level ($E_F$) lies close to lower peak. Below $T_c$ the density of states around at $E_F$ is increased as the redistribution of states occurs due to enhanced hopping. In presence of large strain, a minimum in density of states evolves into a gap, symmetric around $E_d^0 = 0$ and the chemical potential varies little with T. The magnetic field effect (H = 0, 5T) on the resistivity is shown in Fig. 2. A large negative magnetoresistance (MR) is found for the system with smaller strain (Ge=0.15) and for T near to $T_c$. The magnetoresistance remains finite for $T > T_c$ and varies as $H^2$ [Inset of Fig. 2] but the MR effect for the fully spin-ordered state ($T \ll T_c$) is much smaller compared with that around $T_c$. This is due to the fact that the spin-spin correlation for $T \ll T_c$ is nearly independent of magnetic field as it has already reached a value near to its saturation. Close to $T_c$, the magnetic field suppresses spin fluctuation and thereby augments the carrier mobility. These results are in qualitative agreement with the experimental trends[7,8,14,15].

The temperature dependence of the thermopower (S) of the manganite system for different values of strain is shown in Fig.3. Thermopower is positive and reaches a maximum around $T_c$. As the system is more strained, S is higher, consistent with the semiconducting behaviour. At $T<<T_c$ S tends to zero almost linearly. The peak position is dependent on the magnitude of strain and shifts towards lower temperature for increasing strain. In the presence of a magnetic field, S is reduced and the decrease in S is appreciable when T is close to $T_c$. This leads to large negative thermopower around $T_c$ (Fig.4- inset). Magneto-thermopower in the fully ferromagnetic state is much reduced compared to that in paramagnetic state. For the more strained case the influence of the magnetic field is negligible.

## Conclusion:

In this paper, the effect of substitution of rare-earth ion in perovskite manganites on transport properties is examined based on a simple model where the $e_g$ - electron of Mn-ion is coupled to the lattice through the Jahn-Teller mechanism and its motion through lattice is largely modified by the Hund's rule coupling to $t_{2g}$ –spin. It is assumed that the strain due to the mismatch of ionic radii (or in as-deposited condition in thin film) removes the orbital degeneracy of the $e_g$ - level and the hopping of charge carrier in $e_g$-band is modulated by the spin-spin correlation of localized $t_{2g}$ –electrons. Considering the spin-spin correlation in the pair approximation, the resistivity and thermopower values are obtained from the current-current correlation function. The resistive transition to a metallic state, near the ferromagnetic transition temperature $T_c$, is found to be associated with enhanced mobility and the density of states arising from strong spin-spin correlation. The magnitude of the resistive drop decreases as strain increases. For large strain the system remains semiconducting at all temperatures due to opening up an energy gap. A large negative magnetoresistance is found for the system with smaller strain. Magnetoresistance remains finite for $T > T_c$ and varies as $H^2$. Thermopower (S) is positive and exhibits a maximum near $T_c$. A large magneto-thermopower, which is dependent on strain, is found around $T_c$. The Jahn-Teller effect removes the orbital degeneracy of the $e_g$ states and the density of states around the Fermi level goes down making the system more semi-conducting-like in the paramagnetic region. Hund's rule coupling enhances the metallic character by increasing the mobility of the carriers and the density of states at the Fermi level for small hole concentration. When the free energy lowering due to J-T splitting is large compared with that due to Hund's rule exchange interaction, the semiconducting state prevails.

## Acknowledgment:

One of the authors (S. B.) is grateful to CSIR for financial assistance.

**Figure Captions**:

Fig.1  Plot of resistivity vs temperature for different values of Ge. Inset shows the variation of magnetization (M) and spin-spin correlation (<SiSj > )of $t_{2g}$ electrons with temperature.

Fig. 2  Variation of resistivity with temperature for magnetic field H= 0 and 5 Tesla. Inset shows magnetoresistance ($\Delta\rho/\rho_0$) with magnetic field for different temperature.

Fig. 3  Temperature dependence of thermopower for different values of Ge.

Fig. 4  Variation of thermopower with temperature for two magnetic fields. Inset shows the variation of $S/S_0$ with field, for different temperature.

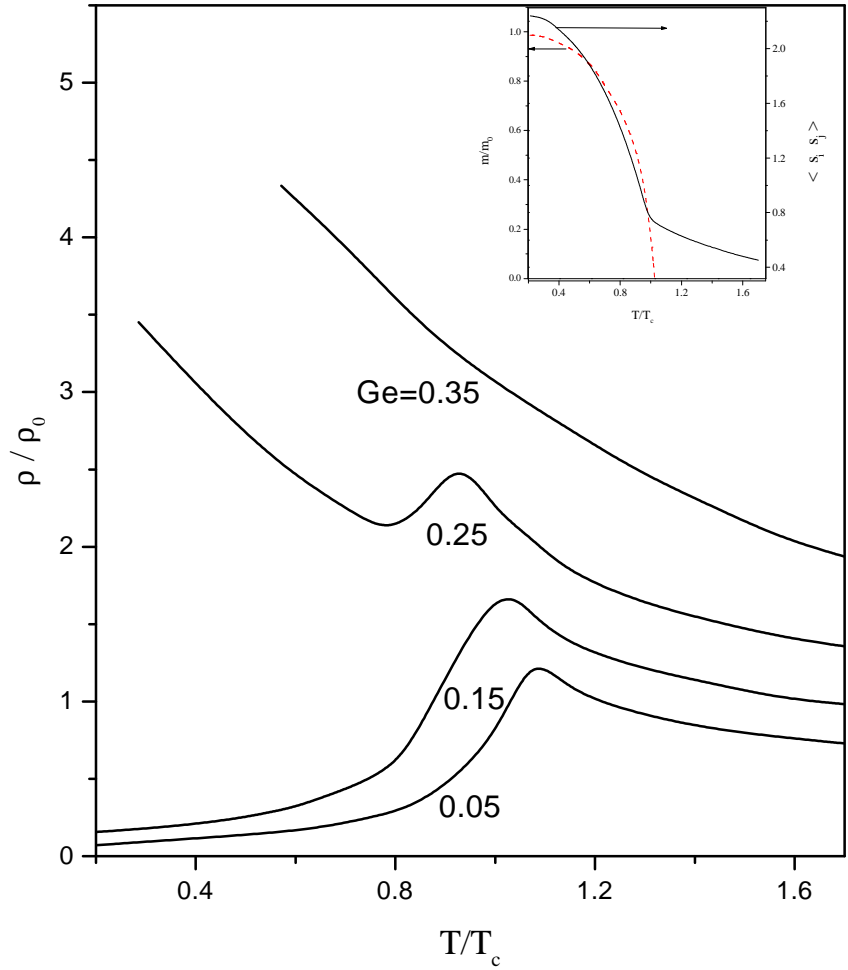

Fig. 1, Basak et al

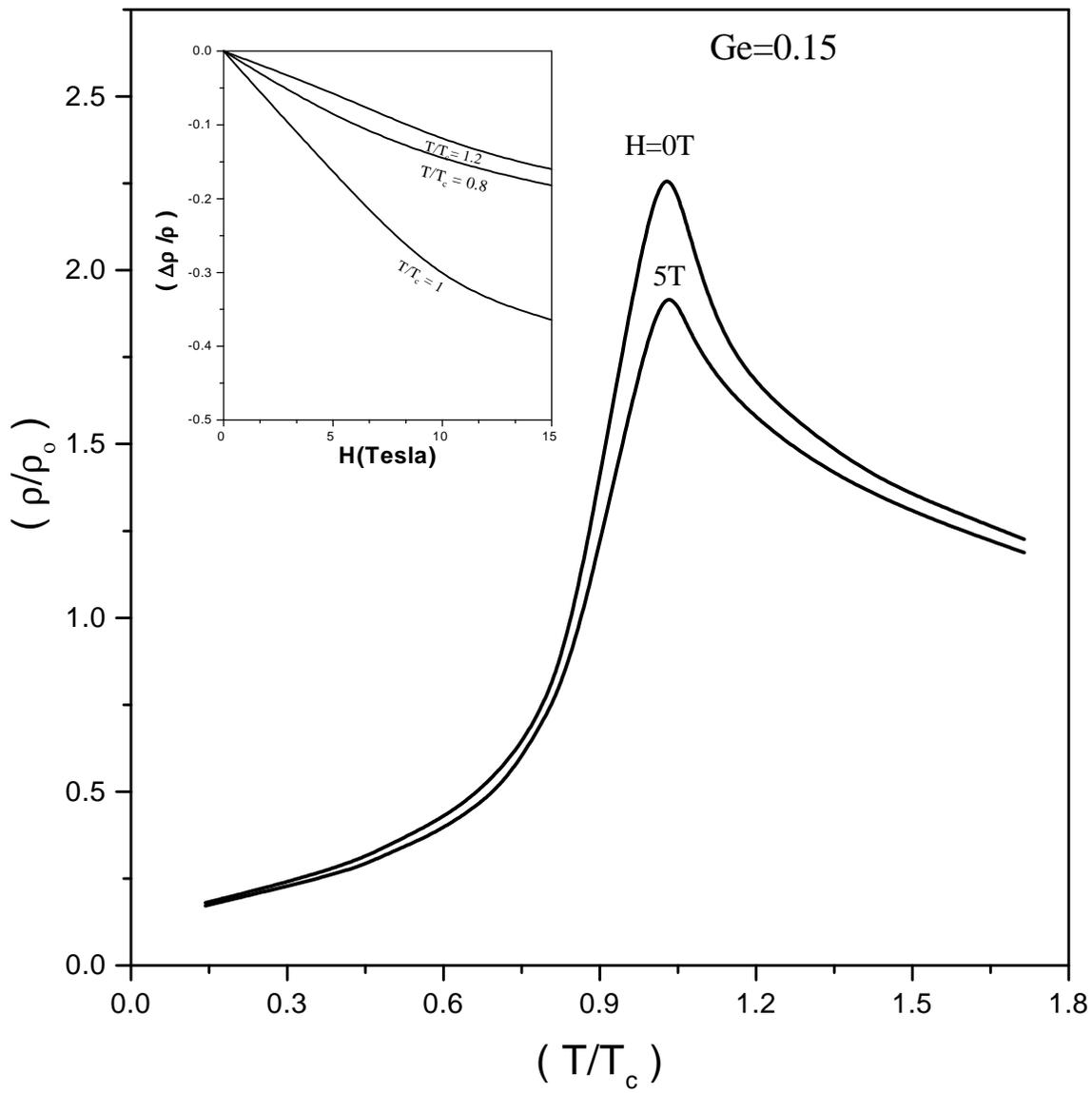

Fig. 2, Basak et al

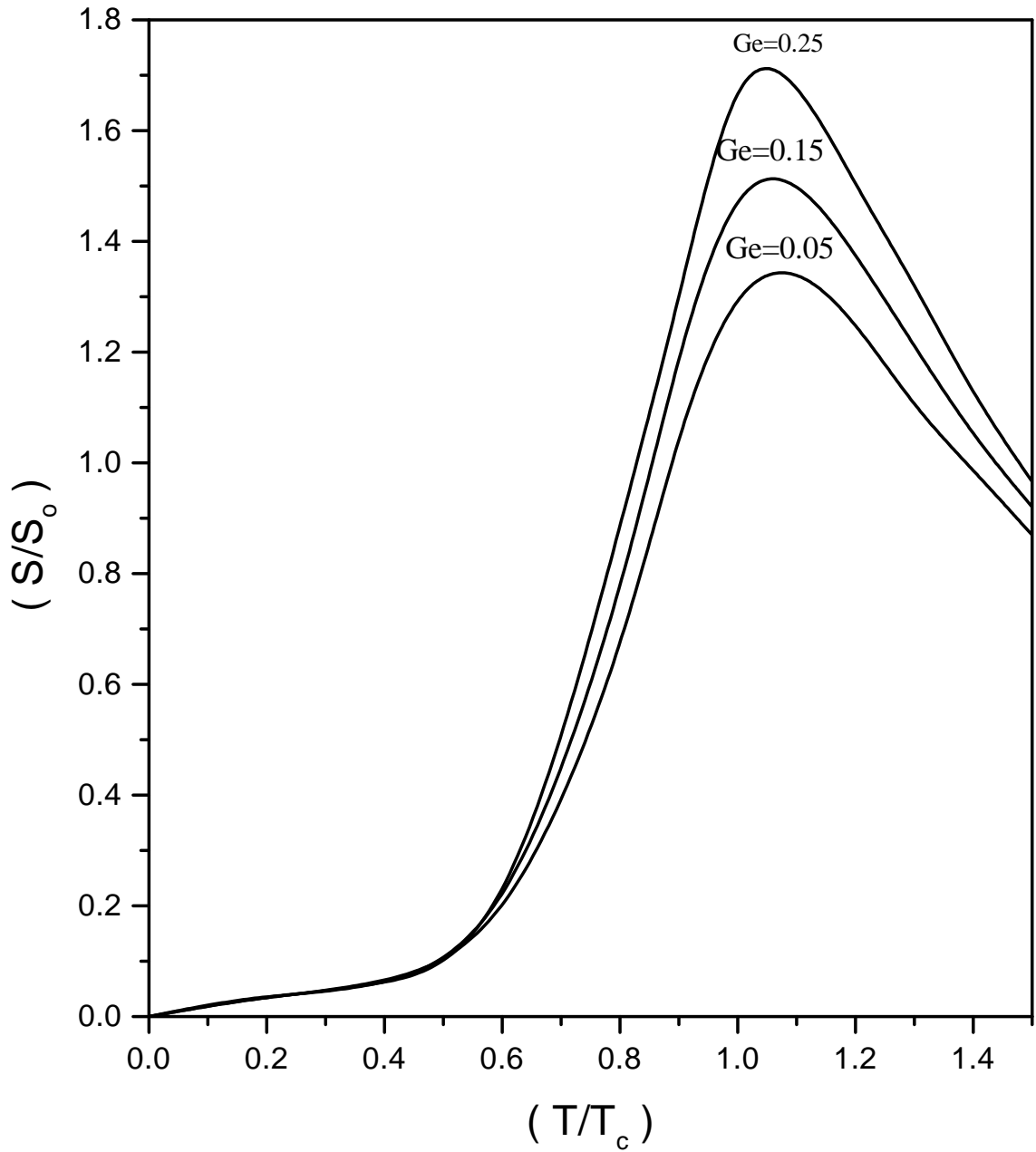

Fig.3, Basak etal

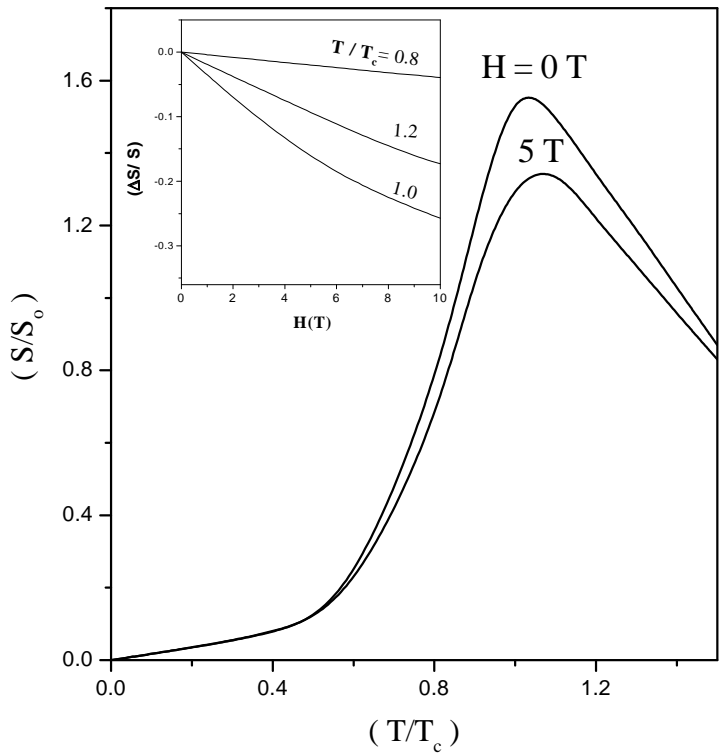

Fig.4, Basak etal